\documentclass{iaus}
\usepackage{graphicx}

\title[Emission line profile Catalog] 
{A Catalog of Emission line profiles for Planetary Nebulae}

\author[C. Morisset and G. Stasinska]   
{Christophe Morisset$^1$ \break \and Grazyna Stasinska$^2$}

\affiliation{$^1$Instituto de Astronomia, UNAM, Mexico 
\break 
email:Morisset@astroscu.UNAM.mx 
\break
$^2$LUTH, Observatoire de Meudon, France}

\pubyear{2006}
\volume{234}  
\pagerange{xxx--xxx}
\date{?? and in revised form ??}
\jname{Proceedings Title IAU Symposium}
\editors{A.C. Editor, B.D. Editor \& C.E. Editor, eds.}
\begin{document}

\maketitle

\begin{abstract}
Using the Cloudy\_3D code (Morisset, this conference),
we run a set of models to build a catalog of emission line profiles of
PNe. The goal is to cover the main morphologies (spherical, ellipsoidal,
bipolar, thin or thick shells) and to compute profiles changing the orientation
of the nebula, the expansion velocity law (including or not
turbulence), the position/size of the aperture. The results is a huge
set of profiles and PV-diagrams that can be compared to observations as a help
to derive morpho-kinematical properties of real objects.
\end{abstract}

\firstsection 


\section{The Catalog}

It is possible to define spherical, ellipsoidal and bipolar
morphologies in a simple way: the inner surface of the nebula is defined as the merge of 2 ellipsoids intersecting in the equatorial plane. Only 2 parameters are then needed: ellipticity
{\it e} and bipolarity {\it b} (distance between the 
lobe centers and the center of the nebula). When {\it b}=0, the nebula is spherical or ellipsoidal. 

We run a set of models to build the catalog. We use 5 shapes (spherical, ellipsoidal, and 3 kind of bipolar, changing the ellipticity of the lobes), 2 shell sizes (thin and thick) and 3 laws for the angular dependence of the inner hydrogen density. These leads to 30 3D-models. We then use 2 laws for the velocity (Hubble flow or constant velocity with turbulence), leading to 60 models with dynamic. Then 3 orientations are used (pole-on, face-on and 45deg), 4 slits are defined (small square, long slit on the major and minor axis and whole nebula) and 2 positions for the slits (centered and slightly  off-center). We compute the line profiles for 5 emission lines (Hbeta, HeI, HeII, [OII] and [OIII]). This leads to 5040 profiles once removing the degenerated ones (rotation of spherical nebulae doesn't change anything).
\section{One page of the Catalog}
The figure shows one of the 144 pages of the current catalog, corresponding to a thick bipolar nebula seen with an angle of 45deg. The first row from top is the description of the nebula (electron density and temperature in the polar and equatorial directions, velocity law and numerical parameters). The following 9 rows are plots for the 5 emission lines, one for each column: the second row shows the surface brightness and the slits superimposed, the third row shows the surface brightness variations among the major and minor axis, the following 4 rows show the line profiles through the slits (bold and light for the center ands off-center apertures resp.), the following two rows the PV-diagrams and the last row the V=0km/s channel map. 

\section{Further developments}

Using the catalog, we will look for all the profiles similar to a given one (using a Khi2 method), to see how different morphologies/expansion law/turbulence can have the same spectroscopic ''look''.
We plan to extend the present catalog to include more morphologies (e.g. open
bipolar nebulae, point-symmetric objects).
The catalog will be linked to the observational catalog from A. Lopez and M.
Richer (this conference).
The catalog/spectra/PV-diags will be made public in electronic form with database queries on the various parameters. 
\begin{figure}
 \includegraphics[height=15cm,width=13cm]{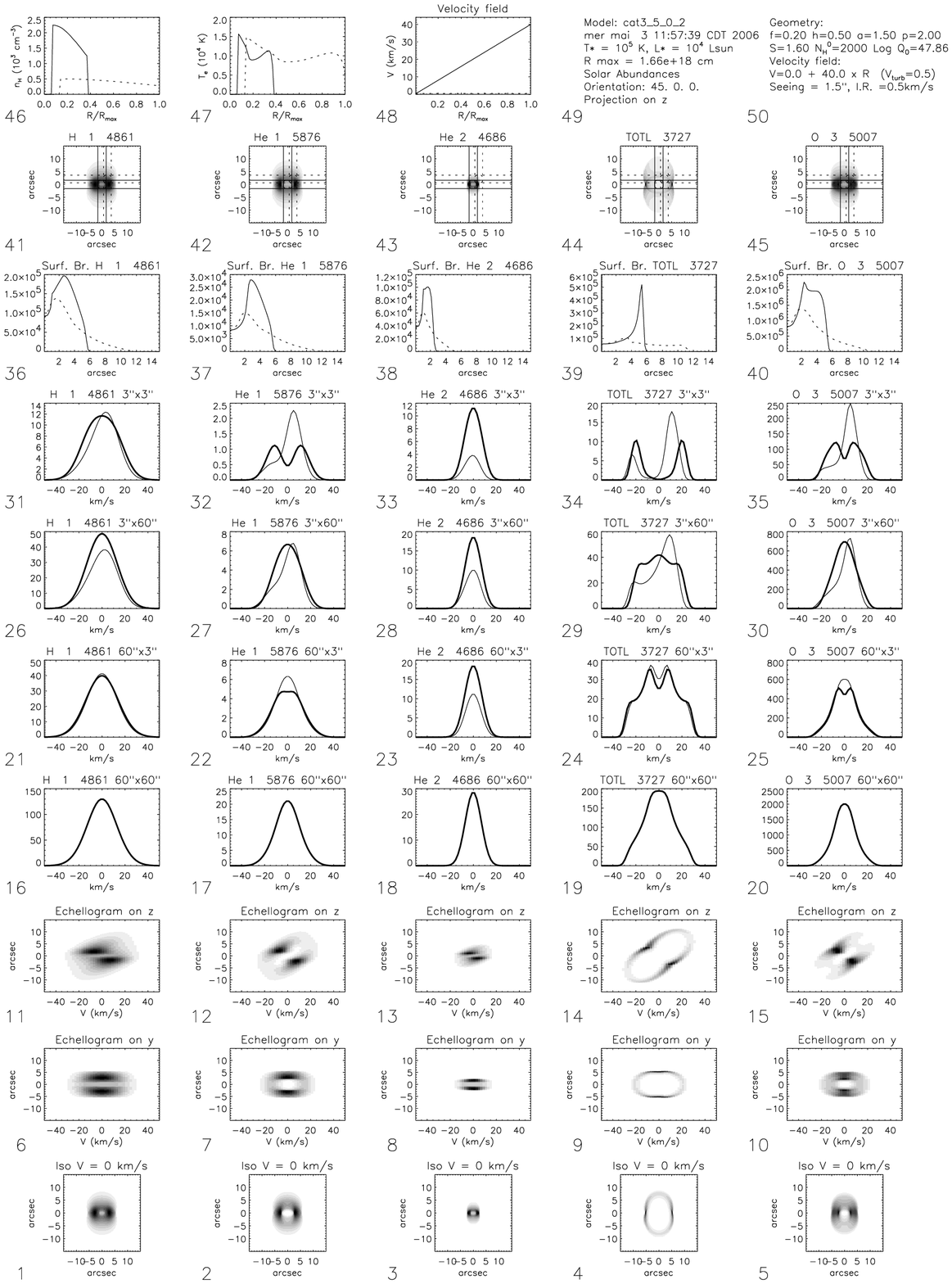}
\end{figure}
\begin{acknowledgments}
The computations are made on a AMD-64bit computer financed by grant PAPIIT IX125304 from DGAPA (UNAM, Mexico). C.M. is partly supported by grant Conacyt-40095 (Mexico).
\end{acknowledgments}

\end{document}